# An Augmented Reality Application and User Study for Understanding and Learning Spatial Transformation Matrices

Zohreh Shaghaghian[1], Heather Burte[2], Dezhen Song[3], and Wei Yan[1]
[1]Texas A&M University, Department of Architecture, College Station, TX 77840, USA
[2]Texas A&M University, Department of Psychological & Brain Sciences
[3]Texas A&M University, Department of Computer Science and Engineering
shaghaghian.zohreh@gmail.com
wyan@tamu.edu

Understanding spatial transformations and their mathematical representations is essential in computer-aided design, robotics, etc. This research has developed and tested an Augmented Reality (AR) application (BRICKxAR/T) to enhance students' learning of spatial transformation matrices. BRICKxAR/T leverages AR features, including information augmentation, physical-virtual object interplay, and embodied learning, to create a novel and effective visualization experience for learning. BRICKxAR/T has been evaluated as a learning intervention using LEGO models as example physical and virtual manipulatives in a user study to assess students' learning gains. The study compared AR (N=29) vs. non-AR (N=30) learning workshops with pre- and post-tests on Purdue Visualization of Rotations Test and math questions. Students' math scores significantly improved after participating in both workshops with the AR workshop tending to show greater improvements. The post-workshop survey showed students were inclined to think BRICKxAR/T an interesting and useful application, and they spent more time learning in AR than non-AR.

## 1 INTRODUCTION AND BACKGROUND

Spatial and mathematical thinking are closely allied. Understanding tightly coupled spatial transformations and mathematical concepts significantly contributes to STEM (Science, Technology, Engineering, and Mathematics) learning in fields of geometric modeling, computer graphics, computer-aided design (CAD), computer vision, robotics, video games, quantum mechanics, and more; and helps students consider mathematics as an interconnected discipline, assisting in higher-level reasoning activities [1]. Literature has clearly acknowledged the difficulty of learning and teaching spatial transformations and the associated math representations such as matrices [1], [2]. Despite the development of Computer Assisted Learning (CAL) in geometry, spatial transformations, and related mathematics, many students still face challenges in solving geometric problems and rely on a trial-and-error process [20].

The close interrelation between math problem solving and spatial thinking has been reported in many studies [3]–[5]. Also, physical model interactions have shown a significant impact on spatial visualization and reduction of extraneous cognitive load in learning spatial and geometry-based problems [6]–[8]. Augmented Reality (AR) as a mediator tool with the ability to superimpose digital content and information over a physical environment supports a context to integrate embodied learning and virtual augmentation of abstract information. AR immersive environment may allow students from different fields of STEM to learn the mathematical logics and abstractions pertained to 3D modeling functions through spatiotemporal experiments while removing the extraneous cognitive load of using keyboards and mice.

In this research, the inherent capability of AR is adopted to realize novel spatial strategies for understanding mathematical concepts through visualizing the concepts and graphical information that are registered (aligned) and synchronized with physical motions in the physical environment. Computer graphics (e.g., arrows, tags, highlighting, etc.) matched with the user's view could effectively draw students' attention [9] and improve their mental imagery. The developed AR app - BRICKxAR/T - realizes the synchronized visualization of mathematical concepts with the physical actions assists students in perceiving a transformation scenario (motions and the corresponding mathematical functions) in one comprehensive experiment. BRICKxAR/T has been evaluated as a learning intervention using LEGO models as example of physical and virtual manipulatives in a user study to assess students' learning gains. We conducted test cases to compare students' learning gain in spatial visualization and math skills in AR and non-AR workshops ($N_{AR} = 29$, $N_{non-AR} = 30$) through pre- and post-tests. Students' math scores significantly improved after participating in both workshops with a tendency towards more improvements in the AR workshop. The post-workshop survey showed students were inclined to think BRICKxAR/T was an interesting and useful application, and they spent more time learning in AR compared to non-AR workshops.

## 1.1 AR for Learning Geometry and Mathematics

Several AR applications have been developed for learning descriptive geometry and mathematics [10]–[13], which demonstrate the positive impact of AR intervention in geometry perception. Most of the apps have used AR applications as visualization tools, displaying 3D geometries and different representations (for example, images of unfolded geometry) in a spatial environment to help students' spatial visualization skills [12]–[17]. The mathematical representations in Construct 3D app are limited to certain graphics only [18] and the mathematical representations in the AR app developed by Cahyono et al., 2018 are static text with no real-time interactions [19]. GeoGebra AR, a major AR application in learning geometry and mathematics [11], has demonstrated positive and significant impact on improving spatial visualization and learning mathematics [20]–[22]. However, GeoGebra AR does not provide any major physical interaction or interplay of physical and digital environments in the learning process. Also, the mathematical section in GeoGebra AR requires advanced pre-knowledge of mathematical equations to generate forms in AR; otherwise, it will be a trial-and-error process. Hence, it maybe a challenging application for a self-learning process. Furthermore, the relation between the mathematics and the corresponding geometry in GeoGebra AR is only one-way; meaning that the user needs to insert a mathematical equation to see the corresponding form or transformations but not the opposite bi-directionally. To the best of the authors' knowledge, no app has developed for learning spatial transformations and the related mathematics.

## 1.2 AR vs. CAD and Virtual Reality (VR)

While CAD and VR may also assist in learning similar spatial and math concepts, compared to AR, they have major limitations. For example, the 2D images of 3D objects in CAD models do not match the real-world perspective view and require a mental model alignment as well as appropriate use of a keyboard and mouse to navigate through the scene. VR can specifically be helpful for simulating experiences that do not exist in the real world (e.g., a fictional environment) or are not easily accessible (e.g., walking through a space) [23], [24]. However, the VR experiments detach a user from the physical environment by replacing it with a complete virtual surrounding [25]. Modeling a whole new environment to simulate a real experience for VR application could be time-consuming and computationally expensive [26]. The interactions in VR environment, realized through external hardware controllers, are neither immediate nor natural, which may impose extraneous cognitive loads. Also, some VR experiments report health issues such as motion sickness and injuries due to wearing VR headsets such as Oculus [23][27]. In contrast, AR supports automatic perspective view alignment with the user's relative position, embodied learning, and physical interactions, which facilitate tangible manipulatives in 3D space, and contribute to improving mental 3D visualization [7], revising mental model misconceptions [28], enhancing spatial cognition and design creativity [6][29], improving idea generation [28][6], and encouraging epistemic action and memory retrieval [30].

## 2   PROTOTYPE DEVELOPMENT AND INTERACTIONS IN THE AR SCENE

The BRICKxAR/T app [31], [32] is an AR prototype for learning spatial transformations and matrix algebra. An equivalent non-AR version of the app is also developed with the same visualization functions without AR features. BRICKxAR/T and the non-AR app are used for conducting a user study with test cases for AR vs. non-AR groups. The app has been developed based on the progressive learning method introduced in literature for learning spatial transformations [33]. In this technique, students will learn spatial transformations in three levels of *motion*, *mapping* and *function* [33]. We have leveraged the AR technology to realize this process for spatiotemporal experiments. Using AR intervention, physical interaction (motion), along with physical and virtual models' interplay (mapping), is supported in the application. Graphical illustrations are displayed in the AR 3D environment, representing the mapping operations. Mathematical functions are synchronized with motion and mapping.

At the start of the user experiment, the app registers two virtual models, one visible and one hidden, through the image marker and aligned with the physical model. When the physical model is moved by the user, the visible virtual model stays in the original location, representing the ***pre-image*** of the transformation, and the hidden virtual model with a visible coordinate system follows the physical model, together representing the ***image*** of the transformation (Figure 1). The physical model's transformations (translation and rotation) are thus visualized by the registered and synchronized distance line and rotation angle graphics as well as the transformation matrix functions (top).

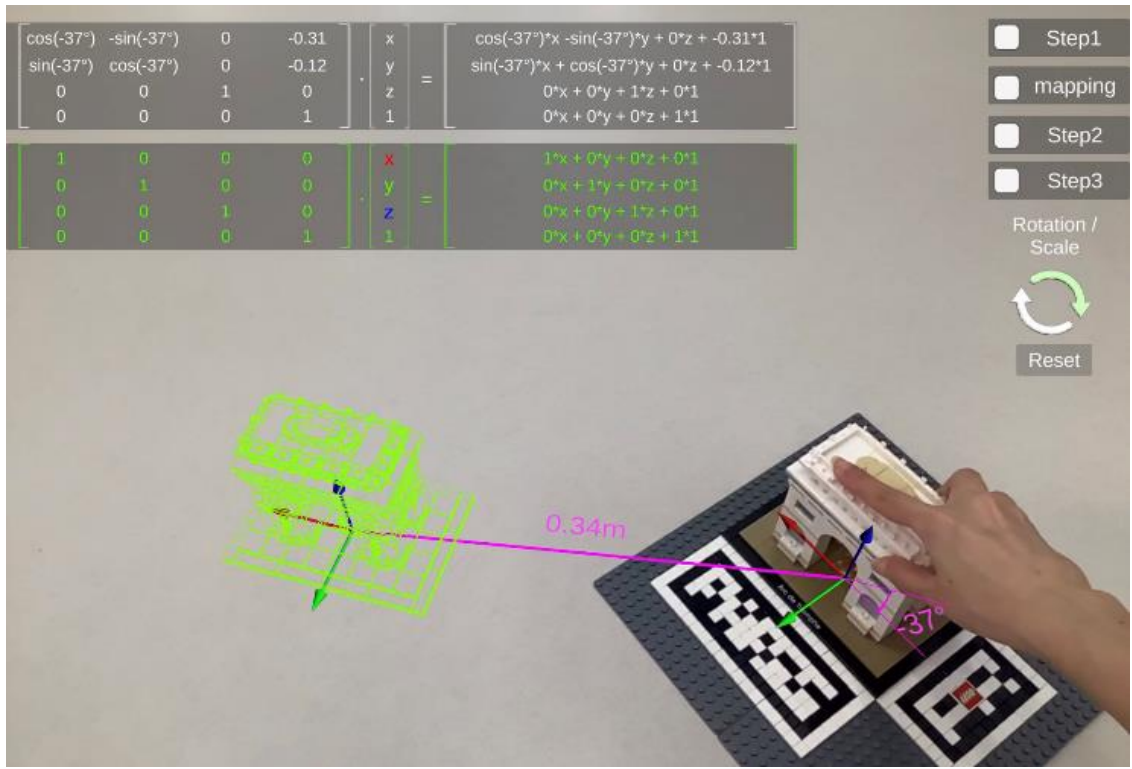

Figure 1: Seen through an AR-enabled iPad, virtual model (pre-image of transformations) and physical model (image of transformations) in the AR scene, with the transformations (translation and rotation) illustrated by synchronized distance line and rotation angle graphics as well as transformation matrix functions (top)

The BRICKxAR/T app has been developed on the AR-enabled iOS mobile device (iPads). In BRICKxAR/T, the AR registration (alignment between virtual and physical models) is highly accurate, and the occlusions between the physical and virtual models are created to realistically reveal the spatial relationships among the models [34]. Hand occlusion is also activated to the AR environment to augment the virtual renderings with correct depth perception with respect to the physical objects (Figure 2).

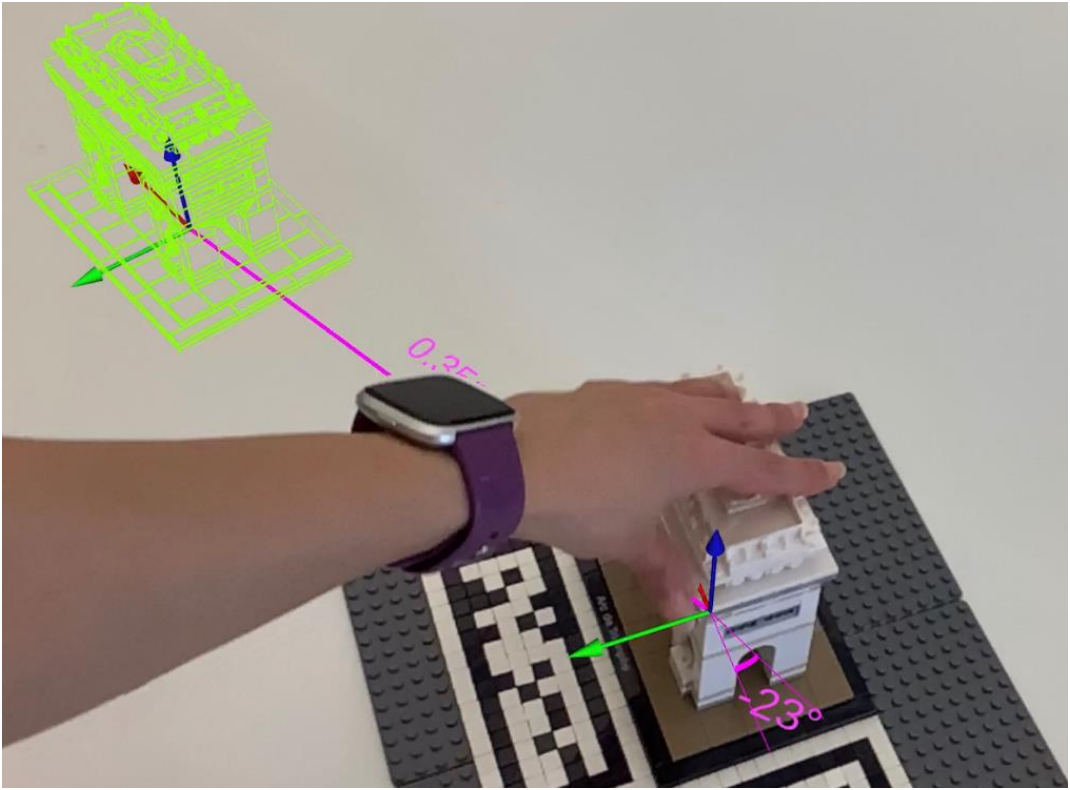

Figure 2: A zoom-in view of the distance line and rotation arc and angle between the physical and virtual models. Occlusion is applied between the real hand and virtual objects

Figure 3 illustrates that the visible virtual model can also be moved through the user interface menus. The 3D Cartesian coordinate system (#1) is the World Coordinate System (WCS) in the AR scene. The WCS gets instantiated in the beginning but never updated during the play. The 3D Cartesian coordinate system (#2) represents the local coordinate system of the physical LEGO model (and an aligned hidden virtual model). This coordinate system gets updated seamlessly by the AR camera tracking the attached image marker and follows the movement of the physical model applied by the user. The wireframe virtual model (#3) can be transformed (translated, rotated, and scaled) through its parameter controls on the AR screen by the user.

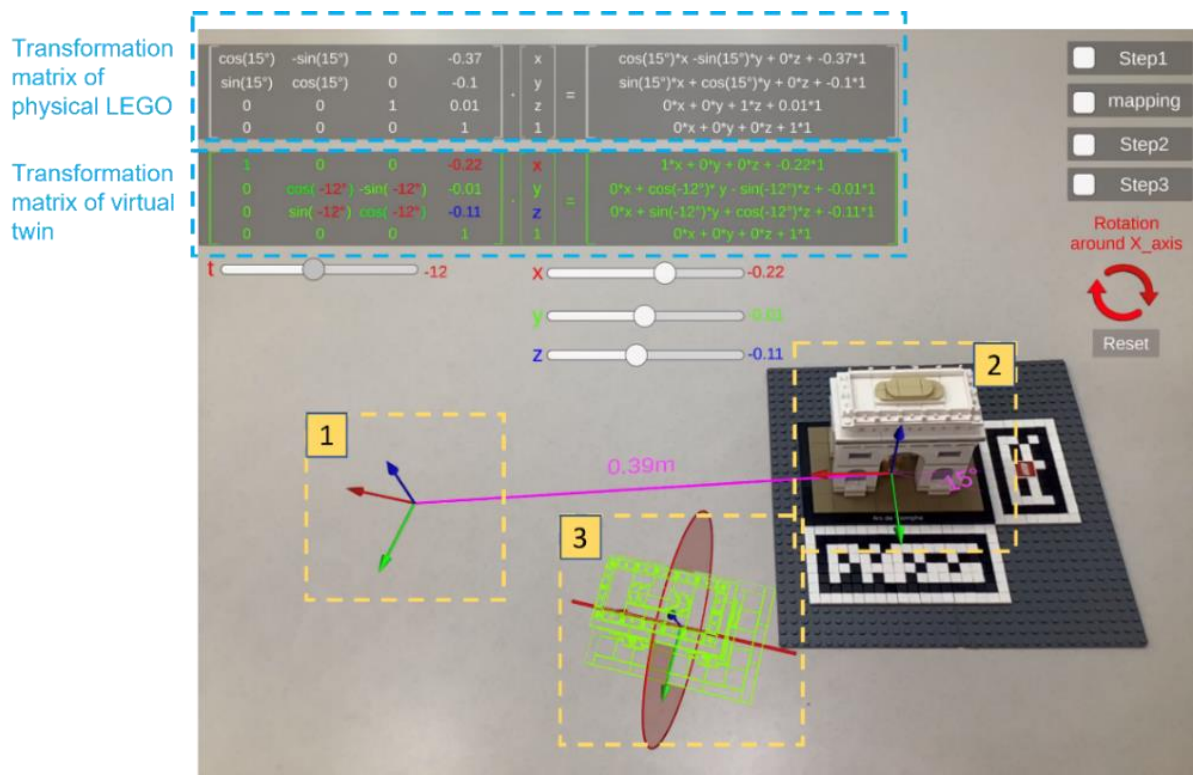

Figure 3: AR scene: physical and virtual models and their corresponding transformation matrices for both models (graphics and notations in yellow and cyan colors are added on top of the AR scene manually for better explanation)

Figure 4 depicts the relations among the AR camera and the three components: physical model, virtual model, and the WCS.

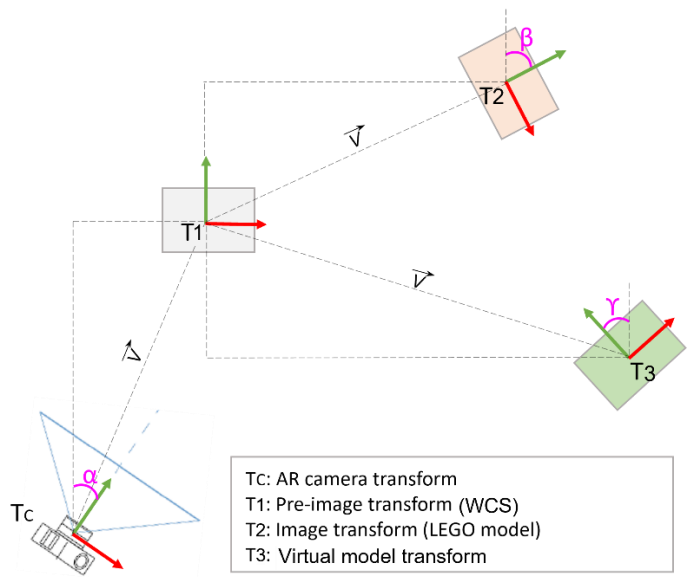

Figure 4: Relations among the AR camera and the three components: physical model, virtual model, and the WCS in the AR scene

The prototype is an RTS (Rotation, Translation, and Scale) puzzle game that helps students understand mathematical notions of spatial transformations along with the mathematical components of transformations, such as variables, parameters, and functions [31]. The mathematical functions matched with the actual motions in real-time assist students to perceive the geometric reasonings behind transformation matrices in an integrated spatial scenario. Students can play with the physical LEGO model, translate and rotate it and observe the corresponding translation and rotation matrices (first row matrices in Figure 3). The distance line and the rotation arc and angle show the graphical representation of the mapping function where the numbers (distance and angle) are matched with the elements in transformation matrices (first row).

Students can also play with the virtual model to numerically compose the transformation matrices (second row) by directly interacting with the corresponding function parameters and associated menu sliders. Translating the virtual model in all three axes can be done at the same time. The associated sliders display when touching x, y, and z parameters of the point vector (representing a point on the geometry) shown in red, green, and blue, respectively (Figure 3).

To rotate or scale the virtual model, students can tap the Rotation/Scale control. Then, the associated transformation matrix and the corresponding slider to change the selected parameter (i.e., rotation angle or scale factor) appears on the screen. The parametric changes applied on the sliders will be reflected in the 4×4 transformation matrix of the virtual model as well as the algebraic equations (results of the matrix multiplication), as shown in Figure 5. To keep the equations of the matrix multiplication simple for the students to understand, rotation around other axes gets zeroed out each time the student chooses a new axis.

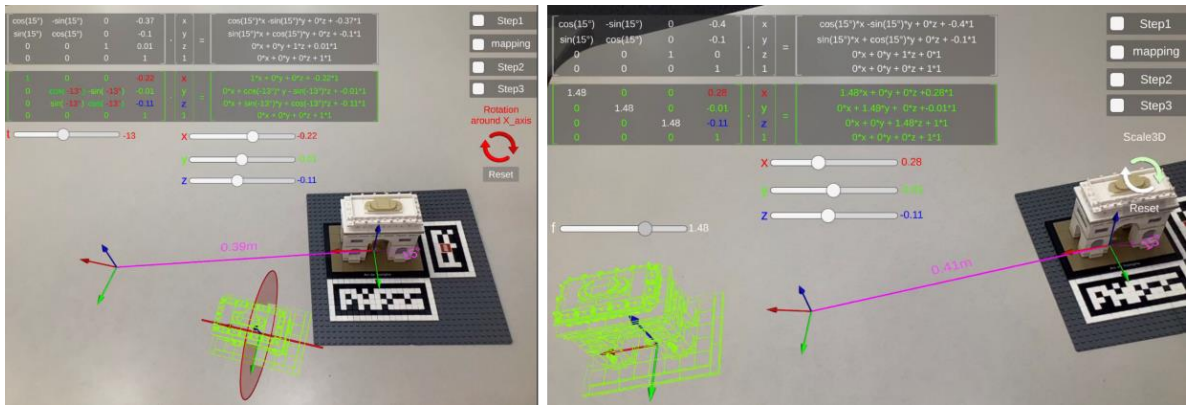

Figure 5: Rotating the virtual model around y-axis (left) and scale in x, y, and z-axes (right)

Finally, students can learn to compose the transformation matrix of the virtual model so that it matches the transformations of the physical model at its current location and orientation (Figure 6). This step intends to intuitively describe the concept behind AR registration (physical and virtual model alignment) as one of the applications of transformation matrices, which is normally done through the AR technology automatically using camera and motion sensors on the AR device.

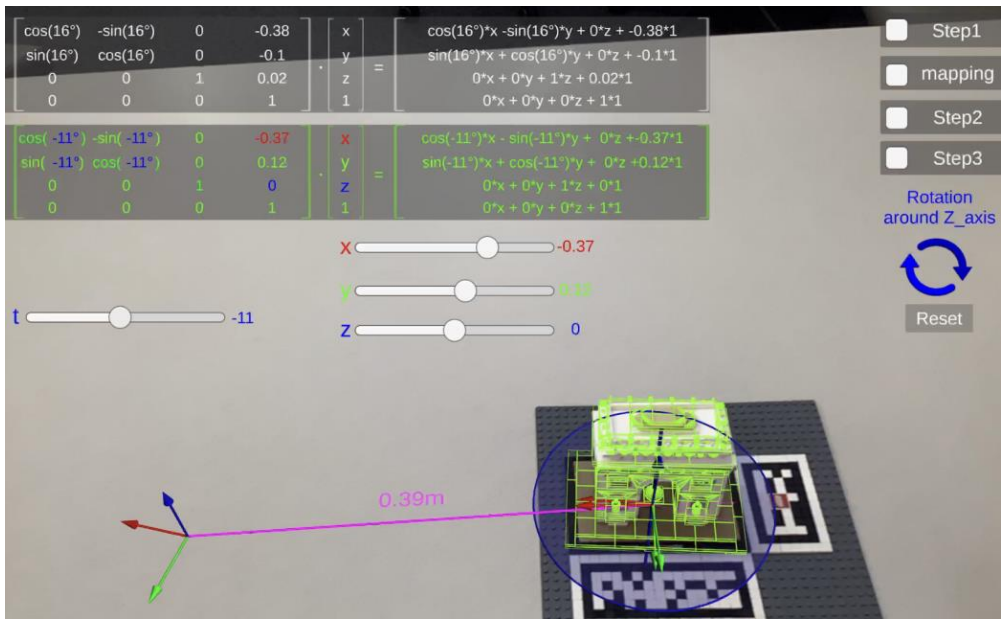

Figure 6: Practicing AR registration through playing with function parameters

## 3 USER STUDY

We conducted test cases to evaluate the apps and compare students' learning gains in math skills between AR and non-AR groups of participants through pre-and post-sessions (Figure 7).

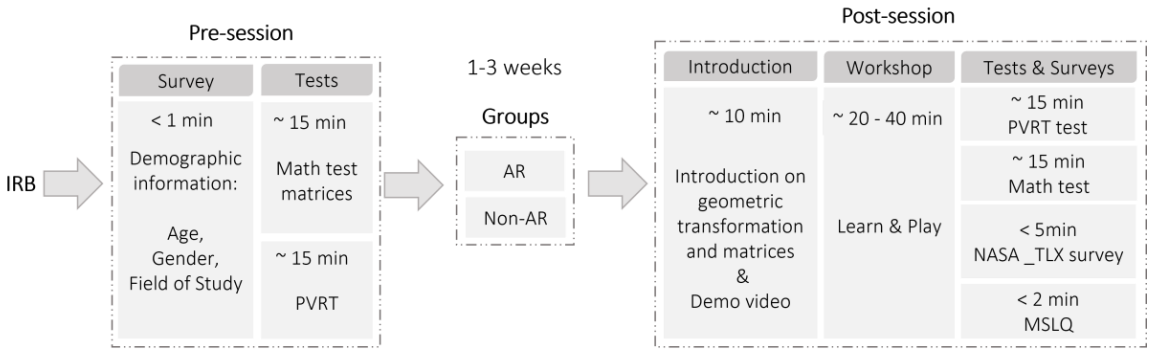

Figure 7: User study framework

Upon the human subject research (Institutional Review Boards, IRB) approval, we invited undergraduate students who had normal vision to use mobile apps through bulk email messages across our university. Participants took an online survey (demographic questionnaire) and pre-tests: Purdue Visualization of Rotation Test (PVRT) and a matrix algebra test. Without disclosing the tests' answers, the students later participated in the (face-to-face) learning workshops and post-tests after a one- to three-week interval to avoid the Testing and History Threat [35]. The order of the questions in the tests remained similar in pre- and post-tests with increased difficulty from the beginning to the end. However, randomization was applied in the answers' choices.

The students in the AR group played and learned through the AR environment integrated with physical models, while the students in the non-AR group played and learned with the non-AR version of the application and physical models separately. During the workshops, each student was provided with an AR-enabled iPad (10.2-inch) with BRICKxAR/T installed, EarPods, and a LEGO set. The LEGO sets were attached to image markers for the AR intervention only. To minimize the "*Cross-Group Contamination*" effects [36], AR and non-AR workshops were held separately and groups were not aware of each other. In the workshops, students first watched an introduction video on the concepts of the learning materials and an instruction demo video on using the corresponding apps (AR or non-AR version). Then, students were asked to play with the apps for 20 to 40 minutes and record their screens during the play. Finally, students took the post-tests (PVRT and matrix algebra test) and surveys (NASA_TLX and Motivated-Strategies-for-Learning-Questionnaire or MSLQ). Students used the provided settings to watch the video lectures, play with the apps, and finally take the post tests.

**3.1 Tests and Surveys**

In order to have an integrated testing platform, all tests and surveys were implemented through the online Qualtrics survey application [37].

*3.1.1 Purdue Visualization of Rotation Test (PVRT)*

We used the short version of PVRT [38] to evaluate students' spatial visualization skills before participating in the workshops to guarantee that students of both groups had similar spatial visualization skills prior to the workshops. Since prior research shows that spatial skills may only improve through repetitive sessions over a long period of time [39], [40], we did not focus on students' PVRT post-test scores and their improvement in a short session workshop. This test consists of 20 multiple choice questions in which students are asked to study how an object is rotated in the sample and select the option that has rotated in the same manner (Figure 8). In our analysis, all questions are equally weighted and the final scores are scaled to 100.

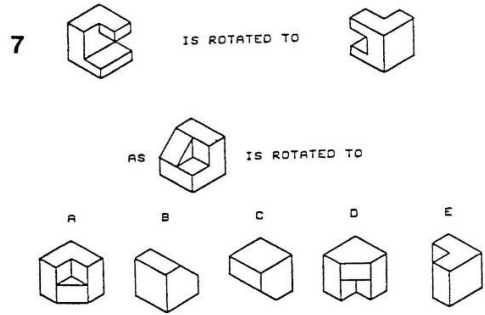

Figure 8: An example question from the 20-item PVRT [38]

*3.1.2 Math Test*

We designed a math test with 13 questions on transformation matrices, based on learning materials of the Khan Academy [41] (Figure 9). In our analysis, all questions are equally weighted and the final scores are scaled to 100.

*Q9.* Which is the best description for the transformation given by the following matrix
$$\begin{pmatrix} 1 & 0 & 0 & 2 \\ 0 & \cos(t) & -\sin(t) & 2 \\ 0 & \sin(t) & \cos(t) & 2 \\ 0 & 0 & 0 & 1 \end{pmatrix}?$$

a) Combined matrix of rotation and move in 3D space

b) Combined matrix of rotation and reflection in 3D space

c) Combined matrix of scale and move in 3D space

d) Combined matrix of scale and rotation in 3D space

Figure 9: An example of the math test designed based on [41]

*3.1.3 NASA_TLX Survey*

We measured the application task workload through the NASA-TLX survey [42] on six dimensions: mental, physical, temporal, effort, frustration, and performance. The survey contains 6 ranking questions in which each dimension is graded on an interval scale ranging from low 0 to high 10, and 15 pairwise questions in which subscale pairs are compared (Figure 10).

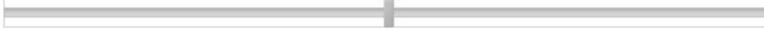

Figure 10: Top: An example of the rating questions; Bottom: an example of the pairwise questions; adopted from NASA_TLX [42]

Based on [42], the adjusted score of each factor is calculated based on its ranking (0 to 100) and its weight (the number of times that the factor is selected in the pairwise questions:

$$Adjusted\ score\ per\ factor = \frac{rating\ score \times weight}{15} \qquad (1)$$

The maximum rating score is 100, while the maximum time that a factor can ever be selected (weight) is 5; hence, the maximum adjusted score that a factor may achieve is 33.3 based on Equation (1).

*3.1.4 Motivated-Strategies-for-Learning-Questionnaire (MSLQ)*

We assessed students' motivations and interests to play with the app through a questionnaire based on the MSLQ survey [43] in three categories of intrinsic value, task anxiety, and self-regulated-learning. Each category consists of multiple questions to evaluate students' subjective viewpoints regarding the corresponding item. The questions are scaled in 5 steps starting from 1 to 5 representing "strongly agree" to "strongly disagree". The mean scores of questions within each category are used in the analyses of this study. Figure 11 shows three example questions from the intrinsic value category.

| | strongly agree | somewhat agree | neither agree or disagree | somewhat disagree | strongly disagree |
|---|---|---|---|---|---|
| I like what I learnt and I think the subject is interesting. | ○ | ○ | ○ | ○ | ○ |
| I think that what I learnt in this workshop is useful for me to know. | ○ | ○ | ○ | ○ | ○ |
| Understanding this subject is important to me. | ○ | ○ | ○ | ○ | ○ |

Figure 11: Questions of the intrinsic value of the motivation questionnaire inspired by MSLQ [9]

## 4 TEST CASES

The AR and non-AR workshops and the tests were held between Fall 2021 and Spring 2022. Twenty-nine (29) students participated in the AR workshop (Figure 12), among which 31% were either very familiar or moderately familiar, and 69% were either slightly or not familiar with digital modeling. 34% of the students were either very familiar or moderately familiar, and 66% were either slightly or not familiar with the AR technology.

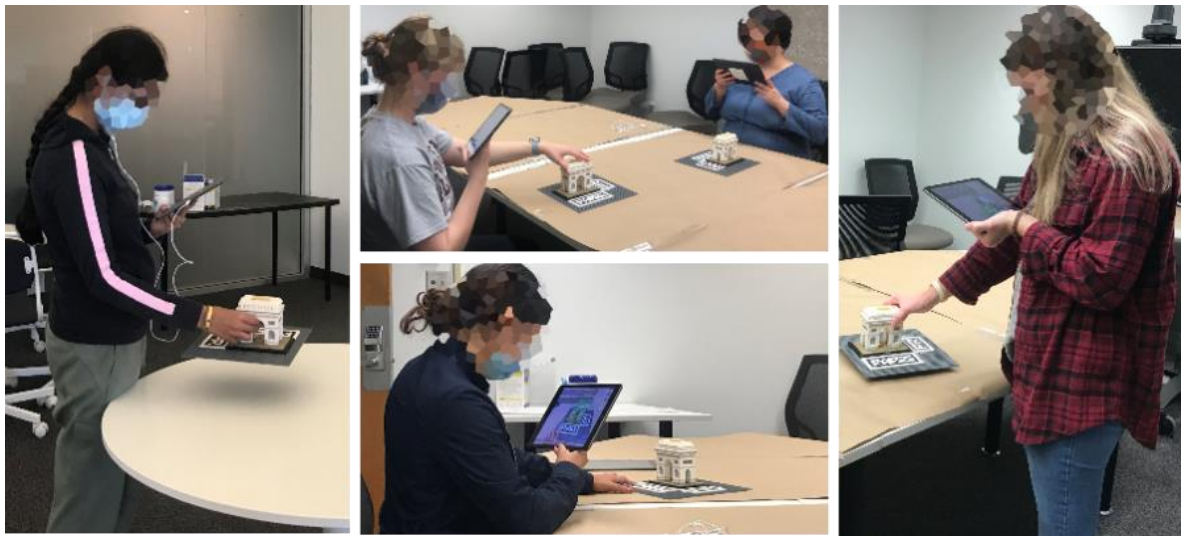

Figure 12: Students playing with the AR version of the app

Thirty (30) students participated in the non-AR workshop (Figure 13), among which 59% were either very familiar or moderately familiar and 41% were either slightly or not familiar with digital modeling (familiarity with AR technology was not asked from the non-AR group).

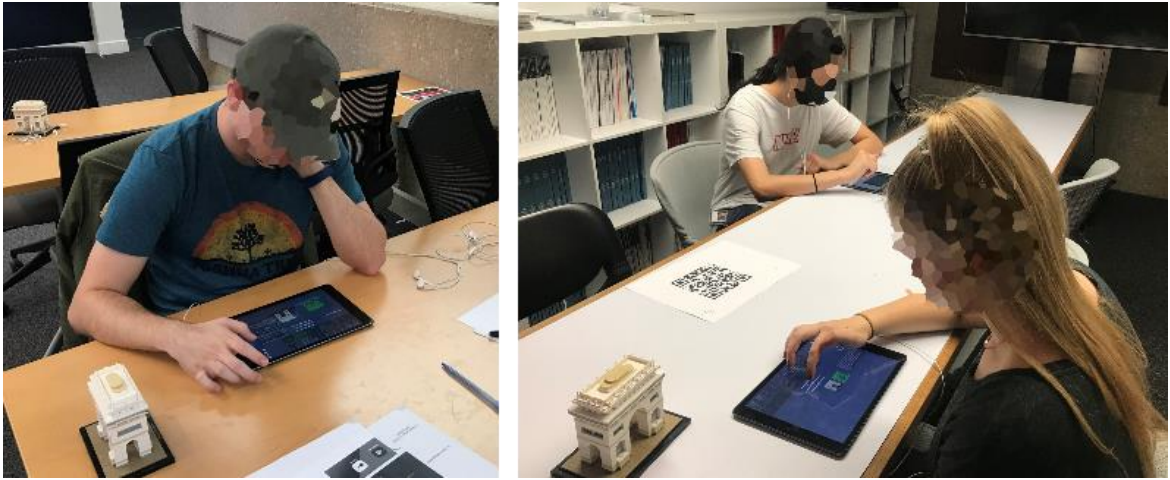

Figure 13: Students playing with the non-AR version of the app

Table 1 shows that female and males students participated in the AR workshop have a ratio of ~ 3 to 1. About half (51.99%) of the AR participants were from STEM colleges (24.14% from engineering and 27.59% from science), and about half (48.27%) from other colleges: architecture (10.34%) or other non-STEM fields (37.93%). In the non-AR workshop, female and male participated with a ratio of ~ 1 to 1. Most participants (73.33%) of the non-AR were from STEM (50% from engineering and 23.33% from science), and only 26.66% were other colleges: architecture (3.33%) or other non-STEM fields (23.33%).

Table 1: Demographic information of the participant of AR and non-AR workshops

| Group | Number | Gender (%) | | | Field of Study (%) | | | |
|---|---|---|---|---|---|---|---|---|
| | | Female | Male | Non-Binary | Engineering | Science | Architecture | Other |
| **AR** | 29 | 72.41 | 24.14 | 3.45 | 24.14 | 27.59 | 10.34 | 37.93 |
| **Non-AR** | 30 | 46.67 | 53.33 | 0 | 50 | 23.33 | 3.33 | 23.33 |

## 5 EVALUATIONS

Statistical tests, such as paired t-tests, ANOVA, and ANCOVA, were conducted on the students' tests and survey results between pre- and post-sessions within each group (AR and non-AR) and between groups (AR vs. non-AR). Before conducting each statistical test, we removed the outliers and validated the corresponding pre-assumptions to determine whether to apply parametric or the corresponding non-parametric tests. We set 0.05 as a threshold for the p-value to accept or reject the null hypothesis associated to each test. Note that the data illustrated in all tables of this section reflect descriptive statistics before removing the outliers while statistical tests were conducted after removing the outliers.

### 5.1 PVRT Scores

Table 2 shows the pre-test PVRT mean scores of the AR and non-AR groups.

Table 2: PVRT pre-test scores

| Group | AR | Non-AR |
|---|---|---|
|  | Pre-test score (%) | Pre-test score (%) |
| **Min** | 20 | 20 |
| **Max** | 100 | 100 |
| **Mean** | 64.14 | 65 |
| **Std** | 23.64 | 23.27 |

The between-subjects ANOVA conducted on the PVRT pre-test scores did not show a significant difference between the AR and non-AR groups ($p = 0.8$), which indicates that students of the two groups had similar mental visualization skills before participating in the workshops for learning spatial transformation matrices.

### 5.2 Math Scores

Table 3 shows that the math mean scores of the AR ($\text{mean}_{\text{pre-test}} = 55.17$, $\text{mean}_{\text{post-test}} = 71.35$) and non-AR groups ($\text{mean}_{\text{pre-test}} = 61.54$, $\text{mean}_{\text{post-test}} = 74.36$) have increased after the workshops with more score improvement in the AR group (29.33%) compared to the non-AR group (20.83%).

Table 3: Descriptive statistical information of the math scores

| | Math scores | | | | | |
|---|---|---|---|---|---|---|
| Group | AR | | | Non-AR | | |
| | Pre-test score (%) | Post-test score (%) | Score improvement (%) | Pre-test score (%) | Post-test score (%) | Score improvement (%) |
| **Min** | 23.1 | 38.5 | +66.7 | 23.1 | 15.4 | -33.3 |
| **Max** | 100 | 100 | 0 | 100 | 100 | 0 |
| **Mean** | 55.17 | 71.35 | +29.33 | 61.54 | 74.36 | +20.83 |
| **Std** | 20.15 | 20.44 | N/A | 20.15 | 23.67 | N/A |

The between-subjects ANOVA showed a near statistically significant difference on students' math pre-test scores between AR vs. non-AR groups ($p = 0.06$, with non-AR scored higher), which means that the non-AR group had better knowledge on the tested subject before the workshops.

The paired t-test showed that students' math scores significantly improved in the post-sessions of AR ($p < 0.001$) and non-AR ($p < 0.001$). It is worth mentioning that both versions had novel visualization features in learning the subject and may be considered as state-of-the-art. Then, we may conclude that visualizing mathematical representations of spatial transformations close to the geometric motions in a "learning-through-play" environment greatly improves students' understanding of the learning materials.

We were especially interested in understanding if the group (AR vs. non-AR) was effective in students' math score improvement. Hence, two ANCOVA tests were conducted for further analysis as follows.

1. ANCOVA on math post-test, with group as the independent variable

This model examined if the post-test score was a function of the pre-test score and whether the function changed based on the group (AR vs. non-AR). The results from the effect test of the model show that the math pre-test is significantly correlated with the math post-test (p < 0.001); however, group (AR vs. non-AR) does not have a significant impact (p = 0.46) on the post-test scores considering the pre-test as the covariate (Table 4). The ANCOVA model with an adjusted R-squared of 0.4026 revealed that 40.26% of the variability of the response variable, i.e., math post-test scores, can be explained by the linear model fitted to the data.

Table 4: Effect test result of ANCOVA on post-test scores having pre-test as the covariate and group as the independent variable

| Variables | F Ratio | p-value |
|---|---|---|
| Math pre-test | 40.06 | <0.001 |
| Group (AR vs. non-AR) | 0.56 | 0.46 |

2. ANCOVA on math post-test, with group as the independent variable and field of study and gender as the extraneous variables

This model examined if the post-test score was a function of the pre-test score and whether the function changed based on the group (AR vs. non-AR), considering field of study and gender as the extraneous variables. Table 5 shows that pre-test is highly correlated with the post-test score (as expected), while neither group, field of study, nor gender has a statistically significant impact on the post-test scores. The summary of fit for this ANCOVA model showed adjusted R-squared equal to 0.421, meaning that 42.1% of the variability of the response variable, i.e., post-test scores, can be explained by this model.

Table 5: Effect test result of the ANCOVA model on post-test scores having pre-test as the covariate, with group, field of study, and gender as independent variables.

| Variables | F Ratio | p-value |
|---|---|---|
| Math pre-test | 25.56 | <0.001 |
| Group | 1.89 | 0.18 |
| Field of study | 0.99 | 0.4 |
| Gender | 0.09 | 0.77 |

### 5.3 Task Load Analysis

We compared the task load of the AR vs. non-AR versions of the app through the NASA_TLX survey. Table 6 shows that students generally rated the prototypes as a low demanding task as the means of the adjusted scores associated with all factors were less than 50% of the maximum possible value of 33.3 (see Equation 1).

Table 6: Mean value of adjusted scores per factor of NASA_TLX results between AR and non-AR groups

| Group | Mental demand | Physical demand | Temporal demand | Effort | Frustration | Performance (negative correlation) |
|---|---|---|---|---|---|---|
| AR | 14.4 | 5.4 | 2.4 | 11.8 | 7.4 | 5.6 |
| non-AR | 11.4 | 2.2 | 3.5 | 8.4 | 4.2 | 5.3 |

The results of ANOVA conducted on the adjusted scores corresponding to each factor showed that students adjusted scores were not significantly different in mental demand (p = 0.14), temporal demand (p = 0.45), frustration (p = 0.09), and performance (p = 0.92) between AR and non-AR groups. However, the data was significantly different in physical demand (p = 0.005), effort (p = 0.02). This outcome reveals that students felt a substantially higher task load, specifically regarding physical demand and effort, in the AR group compared to their peers in the non-AR group, which may be derived from the physical load of holding the device and playing with the physical model at the same time (in the AR workshop) and their unfamiliarity with the AR environment.

### 5.4 Motivation Analysis

We evaluated students' motivations in three categories through a motivation questionnaire (adopted from MSLQ [43]) (Table 7).

Table 7: Mean values of students' ratings to the questions of the MSLQ survey (range 1 strongly agree to 5 strongly disagree)

| Category | Item | Mean value AR | Mean value non-AR |
|---|---|---|---|
| Intrinsic-value | I like what I learned, and I think the subject is interesting. | 1.7 | 1.9 |
| Intrinsic-value | I think that what I learned in this workshop is useful for me to know. | 2.1 | 2 |
| Intrinsic-value | Understanding this subject is important to me. | 2.6 | 2.9 |
| Intrinsic-value | Mean of the intrinsic value category questions | 2.1 | 2.3 |
| Task-anxiety | I was so nervous during the workshop that I could not remember the material I had learned. | 3.6 | 3.5 |
| Task-anxiety | I had an uneasy, upset feeling when I was participating in the workshop. | 4.3 | 4.6 |
| Task-anxiety | When I was performing the task in the workshop, I thought about how poorly I was doing. | 3.6 | 3.9 |
| Task-anxiety | Mean of the task-anxiety category questions | 3.8 | 4 |
| Self-regulated-learning | When I was taking the tests (math and mental rotation test), I put together what I learned in the workshop and lecture. | 1.6 | 1.8 |
| Self-regulated-learning | When I was taking the math test, I used visual imagery to visualize the geometric transformations and what I experienced in the workshop. | 1.6 | 1.9 |
| Self-regulated-learning | Visualization of matrix representations during the workshop helped me in solving the math test. | 1.8 | 2.2 |
| Self-regulated-learning | Mean of the self-regulated-learning category questions | 1.7 | 2 |

The result of the non-parametric Wilcoxon Rank Sum tests on students' answers to each category did not show any significant difference between AR vs. non-AR groups ($p_{intrinsic-value}$ = 0.31, $p_{task-anxiety}$ = 0.55, $p_{self-regulated-learning}$ = 0.11). For both groups, the results of the intrinsic-value category show that students agreed (total $mean_{AR}$ = 2.1, total $mean_{non-AR}$ = 2.3) that what they learned in the workshop was an interesting, useful, and important subject to learn, with the AR group agreed more than the non-AR group. The task-anxiety category results reveal that students somewhat disagreed (total

mean$_{AR}$ = 3.8 and total mean$_{non-AR}$ = 4) about how uneasy and upset they felt during the workshop, with the non-AR group disagreed more than the AR group. The results from the self-regulated-learning category show that students agreed (total mean$_{AR}$ = 1.7, total mean$_{non-AR}$ = 2) that the learning materials of the workshops were helpful in self-learning and answering the questions, with AR group agreed more than the non-AR group.

### 5.5 Play Time

We used the data of screen recordings to compare the length of students' play time in the AR vs. non-AR workshops (Table 8).

Table 8: Descriptive statistical information of students play time with the apps

| | **Play time (minute)** | |
| --- | --- | --- |
| **Group** | **AR** | **non-AR** |
| **Min** | 17 | 8.78 |
| **Max** | 35.95 | 22.65 |
| **Mean** | 25.33 | 15.99 |
| **Std** | 5.56 | 3.8 |

The result of the non-parametric Wilcoxon Rank Sum tests on students' play time with the apps showed that students played more with the AR app during the workshops than the non-AR app, with a significant difference ($p < 0.001$). Based on the play time we may interpret that the AR app was more interesting and engaging for the students to learn through play. Our records show that 37.93% of the students in the non-AR group quit playing in less than 15 minutes and only 31.03% of the students played 20 minutes or more (only 3.45% played 21 minutes or more). On the other hand, all students of AR group played more than 15 minutes while 89.66 % of the students played 20 minutes or more (68.97% played 21 minutes or more).

### 6 DISCUSSIONS, CONCLUSIONS, AND FUTURE WORK

This paper presented BRICKxAR/T and the user study with test cases to evaluate the apps in the AR vs. non-AR environments. BRICKxAR/T is an AR educational tool for learning the geometric reasoning behind mathematical representations of spatial transformations. In the test cases, we first guaranteed that both groups had similar spatial visualization skills prior to participation in the workshops through PVRT. Then, we assessed students' learning gains in math skills through the math test on transformation matrices before and after participating in the workshops and compared the results between the AR and non-AR groups. The application task load and participants' motivations were also evaluated through NASA_TLX and MSLQ surveys, respectively.

Based on the between-subjects ANOVA result of the PVRT pre-test scores, students had similar spatial visualization skills before participating in the workshops. Based on the between-subjects ANOVA result of the math pre-test scores, students of non-AR group had better knowledge (near statistically significant difference) than the AR group before the workshops. Students' math scores for matrix algebra improved significantly after the workshops in both of the AR and non-AR groups, with more improvements in the AR group. Hence, the researchers conclude that the features integrated in BRICKxAR//T interventions (in both AR and non-AR, and especially in AR) may improve students' understanding of mathematical representations of spatial transformations.

This research has not conducted a comparative study between the AR group and a conventional "control group" (for example, a group that learns the same topic through conventional methods, such as lectures, handbooks, or videos). The reason is that CAL has been practiced widely and has already shown significant improvement compared to the conventional methods due to literature. AR can be considered as a novel CAL method, and thus we decided to compare BRICKxAR/T with a higher bar, which is our non-AR setting, and the results are promising. Especially, students in the AR group agreed more than the non-AR group on that what they learned in the workshop was an interesting, useful, and important subject to learn, and that the learning materials of the workshops were helpful in self-learning and answering the questions. More importantly, the data from the screen recordings showed that students were willing to spend significantly more time playing with AR than non-AR app. This suggests that students are more interested and engaging in the BRICKxAR/T learning environment. This result is also aligned with researchers' observations: most students of AR group were more excited and curious to play with the AR app and spent more time to play and learn, while many students in the non-AR group got bored early and stopped playing and learning.

BRICKxAR/T has the potential to support many students who struggle with spatial and math reasoning, especially those from underrepresented groups in STEM. Gaining a fundamental understanding of spatial transformations can uniquely contribute to students' learning and development of spatial reasoning and allied mathematical skills, leading to improve STEM coursework, STEM retention, and degree attainment, and thus supports students' future development of expertise and career success across STEM disciplines.

The contributions of this study include the following computer-human interaction technology and learning innovations:
- BRICKxAR/T is an innovative learning environment that enables physical and virtual interplays to engage students in embodied learning (by transforming physical models by hand).
- BRICKxAR/T integrates spatial transformation matrices and related math information with the physical model movement controlled by students in AR, making difficult invisible concepts visible for supporting an intuitive and formal understanding of spatial reasoning and mathematical formulation.
- .BRICKxAR/T showed the potential to help students conceive, connect, and compare math conceptions of motions, mappings, and functions in AR to help overcome well-documented difficulties students face when learning spatial transformations and allied mathematical representations.
- The functions of BRICKxAR/T help students see relationships between spatial manipulations and mathematical operations, bridging the spatial-mathematical divide.
- BRICKxAR/T's motion tracking of physical object transformations and the student's hand that controls the objects showed the potential to help collect fine-grained behavior data to enhance learning analytics, for example, what the rotation axes and angles are when a hand rotates a physical model.

The results from the NASA_TLX showed that students perceived significantly more physical load and effort while playing in the AR workshop compared to their peers in the non-AR workshop. The researchers believe that the outcome may be driven by the current limitations of BRICKxAR/T, which require students to hold the iPad (an affordable AR device for learning) with one hand and manipulate the physical model or the screen user interface with the other hand. Also, the unfamiliarity of students with the AR environment may impact their subjective assessment of the effort factor, meaning that they felt more effort in learning the new environment. Leveraging 3D model registration (such as Vuforia Model Target [44]) and exceling the implementation through immersive devices (such as the more convenient and immersive future AR glasses) may improve the task load rating by the AR users. Addressing the limitations of the current application in future work is expected to significantly improve the learning of spatial transformations and their mathematical representations utilizing AR technologies.

**ACKNOWLEDGMENTS**

This material is based upon work supported by the National Science Foundation under Grant No. 2119549.